\documentclass[10pt]{iopart}
\usepackage{graphicx}
\usepackage{iopams}
\usepackage{bm}

\begin{document}

\renewcommand*{\max}{{\mathrm{max}}}
\newcommand*{\eff}{\mathrm{eff}}
\newcommand*{\h}{\mathrm{h}}
\renewcommand*{\b}{\mathrm{b}}
\newcommand*{\cosmo}{\mathrm{m}}
\newcommand*{\di}{\partial}
\newcommand*{\approaches}[2]{\xrightarrow[#2]{\,\,\,{#1}\,\,\,}}

\title{Ricci flows, wormholes and critical phenomena}

\bibliographystyle{iopart-num}

\author{Viqar Husain and Sanjeev S.~Seahra}

\address{Department of Mathematics \& Statistics, University of
New Brunswick, Fredericton, NB E3B 5A3, Canada}

\begin{abstract}
We study the evolution of wormhole geometries under Ricci flow using
numerical methods. Depending on values of initial data parameters,
wormhole throats either pinch off or evolve to a monotonically
growing state. The transition between these two behaviors exhibits a
from of critical phenomena reminiscent of that observed in
gravitational collapse. Similar results are obtained for initial
data that describe space bubbles attached to asymptotically flat
regions. Our numerical methods are applicable to ``matter-coupled''
Ricci flows derived from conformal invariance in string theory.
\end{abstract}

\section{Introduction}

An analog of the diffusion equation for geometries is the so-called
Ricci flow. This is a type of parabolic partial differential
equation that homogenizes geometry in a manner similar to the
homogenization of density or heat produced by the diffusion
equation.

The Ricci flow equation was first studied in the early 1980s.  It
arose nearly simultaneously in mathematics \cite{hamilt}, in the
study of the classification of 3-geometries, and in string theory
physics \cite{friedan,sen} from the requirement that a quantized
string  retain conformal invariance. Certain modifications of the
purely geometric flow equation by ``matter terms'' motivated by
string theory were central to Perelman's \cite{perel} work on the
Poincare conjecture.

A goal of analytic work on the flow equations is to study the
approach to singularities, and to understand the  so-called
``ancient solutions,''  which are the non-linear analogs of the heat
kernel of the diffusion equation. Much of the work in this area has
focused on the flow of compact geometries \cite{cao-chow}, with
relatively little on non-compact cases such as the asymptotically
flat or anti-deSitter geometries. In the asymptotically flat case
for dimension $n\ge3$, it has been demonstrated recently that the
flow exists and remains asymptotically flat for an interval of time,
with the mass remaining constant \cite{ow-asympflat,dai-ma}. For the
rotationally symmetric case, Ref. \cite{ow-asympflat} also shows
that the flow is eternal for data containing no minimal surfaces.

The purpose of this paper is to investigate numerically some
examples of  the flow of asymptotically flat geometries, and to
describe techniques that are easy to generalize to cases where
matter flow equations are included.

Previous numerical work on Ricci  flows focused on the (compact)
case of a 3-sphere that is corseted (i.e., a surface formed by
uniformly shrinking the equator of a 3-sphere), where it was found
that the manifold flows to a round 3-sphere or a neck pinch
depending on the degree of corseting \cite{gi}. The flow of horizon
area and Hawking mass has also been studied \cite{sam} for initial
data that is viewed as time symmetric data for the Einstein
constraint equations. Numerical work pertaining to string theory
aims at obtaining new black hole solutions as fixed points
\cite{hw}.

The focus of the present work is the geometric flow of wormhole
geometries. We study several  classes of initial data  and find two
possible outcomes: the wormhole ``pinches off'' leaving two disjoint
asymptotically flat regions, or it expands indefinitely. We also
find that there is a type of critical behaviour that separates these
extreme cases. The numerical methods we utilize are stable for long
time flows, and have potential applicability to the largely
unexplored geometry-matter cases.

\section{Flow equations}

The Hamilton-DeTurck Ricci flow is defined by the equation
\cite{hamilt,deturck}
\begin{equation}\label{eq:flow equation 1}
    \di_t g_{ab} = -2R_{ab} + 2 \nabla_{(a} V_{b)}.
\end{equation}
For our work, we take $g_{ab} = g_{ab}(t,x^c)$ to be the 3-metric on
a manifold $\Sigma_t$ with Euclidean signature and $R_{ab}$ is the
associated Ricci tensor. The parameter $t$ labels different
3-geometries along the flow and should not be confused with the
intrinsic 3-dimensional coordinates $x^c$. Finally, $V^a$ is a
DeTurck vector field that generates diffeomorphisms along the flow.
The DeTurck field essentially represents the freedom to change
3-dimensional coordinates as the flow progresses.  It is useful to
note that $t$ has dimensions of $[\mathrm{length}]^{2}$ and $V^a$
has dimensions of $[\mathrm{length}]^{-1}$.

We assume the 3-manifolds $\Sigma_t$ are spherically symmetric, for
which  a general metric \emph{ansatz}  is
\begin{equation}\label{eq:metric ansatz}
    ds^2 = e^{2X(t,\rho)}[ d\rho^2 + R^2(t,\rho)d\Omega^2 ].
\end{equation}
Spherical symmetry implies the following form for the DeTurck field:
\begin{equation}\label{eq:vector ansatz}
    V^a = V(t,\rho) \di_\rho.
\end{equation}
When (\ref{eq:metric ansatz}) and (\ref{eq:vector ansatz}) are put
into (\ref{eq:flow equation 1}) we obtain two independent flow
equations for $\di_t X(t,\rho)$ and $\di_t R(t,\rho)$.  There is no
dynamical equation for $V(t,\rho)$, so the system is
underdetermined, which reflects the coordinate freedom embodied by
the DeTurck field.  To close the system, we need to fix coordinate
gauges. There are three choices we have investigated:
\begin{enumerate}
    \item\label{case:1} $V(t,\rho) \equiv 0$;
    \item\label{case:2} $R(t,\rho) \equiv \rho$ (conformally flat gauge); or
    \item\label{case:3} $X(t,\rho) \equiv 0$ (areal radius gauge).
\end{enumerate}
Enforcement of condition (\ref{case:2}) or (\ref{case:3}) in the
flow equations (\ref{eq:flow equation 1}) results in a single
dynamical equation for $X(t,\rho)$ or $R(t,\rho)$, respectively, and
an equation of constraint for $V(t,\rho)$.  The choice
(\ref{case:1}) gives two dynamical equations for both metric
functions.

For our purposes, a wormhole geometry is defined by a metric
(\ref{eq:metric ansatz}) that has two asymptotically flat regions as
$\rho \rightarrow \pm \infty$.  The combination $e^{X(t,\rho)}
R(t,\rho)$ should be non-zero for all $\rho$, and the wormhole
throat (or throats) are located at local minima $\rho =
\rho_\mathrm{th}$ of this function\footnote{If we interpret the
3-geometries $\Sigma_t$ as being embedded in 4-dimensional
Lorentzian spacetimes as moments of time symmetry, each wormhole
throat represents an apparent horizon.  However, if the $\Sigma_t$
are embedded with non-zero extrinsic curvature, such an
interpretation is not possible.}
\begin{equation}
    \di_\rho(e^{X} R)\big|_{\rho = \rho_\mathrm{th}} = 0,
    \quad \di^2_\rho (e^{X} R)\big|_{\rho = \rho_\mathrm{th}} > 0.
\end{equation}
 The definition of the wormhole throat suggests a convenient
coordinate gauge choice is given by case (\ref{case:3}) above, with
line element
\begin{equation}\label{eq:areal radius gauge}
    ds^2 = d\rho^2 + R^2(t,\rho) \, d\Omega^2.
\end{equation}
In this gauge the area of 2-spheres of constant $\rho$ is $4\pi
R^2(t,\rho)$, hence the name ``areal radius'' gauge. The condition
$X \equiv 0$ reduces the flow equations (\ref{eq:flow equation 1})
to
\numparts\label{eq:main equations}
\begin{eqnarray}\label{eq:flow equation 2}
    \di_t R = \di^2_\rho R + \frac{(\di_\rho R)^2}{R} - \frac{1}{R}
    + V \di_\rho R, \\ \label{eq:constraint equation}
    \di_\rho V = -2\frac{\di_\rho^2 R}{R}.
\end{eqnarray}
\endnumparts
As mentioned above, the latter is a constraint equation. These
equations must be supplemented by boundary conditions and initial
data.  In the areal radius gauge, asymptotic flatness requires that
\begin{equation}\label{eq:asymptotic BC 1}
    R \sim |\rho| + \mathcal{O}\left(l_1 \ln \left(|\rho|/l_2\right) \right),
\end{equation}
where $l_1$ and $l_2$ are two length scales. Putting this asymptotic
limit into (\ref{eq:flow equation 2}) and (\ref{eq:constraint
equation}), we obtain
\begin{equation}\label{eq:asymptotic BC 2}
    V \sim \mathcal{O}\left( |\rho|^{-2} \right).
\end{equation}
The asymptotic behaviour of $R$ suggests that we search for
solutions that are  even functions of $\rho$.  We then only need to
consider the interval $\rho \in [0,\infty)$, and we have the
additional boundary condition
\begin{equation}\label{eq:Neumann}
    \di_\rho R \big|_{\rho = 0} = 0.
\end{equation}
The boundary conditions imply that there is always a local extrema
of $R$ at $\rho = 0$. If this is a minima, then there is a throat at
$\rho = 0$.  To complete the specification of the flow, we need to
give initial data for $R$ at $t = 0$.  The various classes of
initial data considered in this work are described in
\S\ref{sec:Morris-Thorne} and \S\ref{sec:bubbles}.

Finally, we note that it is possible to embed the 3-geometries
(\ref{eq:areal radius gauge}) in 4-dimensional flat space
\begin{equation}
    ds^2 = dR^2 + dZ^2 + R^2 d\Omega^2,
\end{equation}
via the parametric equations
\begin{equation}
    R = R(t,\rho), \quad Z = Z(t,\rho), \quad \di_\rho Z =
    \sqrt{1-(\di_\rho R)^2}.
\end{equation}
These embeddings are only possible for areal radius functions
satisfying $|\di_\rho R| < 1$.  We use these as a tool for the
visualization of Ricci flow initial data in Figs. \ref{fig:embedding
1} and \ref{fig:embedding 2}.

\section{Numerical method}\label{sec:numerical}

We solve the equations (\ref{eq:flow equation 2}) and
(\ref{eq:constraint equation}) numerically using finite difference
methods.  The computational domain is defined by $\rho \in
[0,\rho_\mathrm{max}]$ and $t \in [0,t_\mathrm{max}]$.  The spatial
and temporal intervals are discretized into elements of size
$\delta\rho$ and $\delta t$, respectively.  The notation $A_i^j$
refers to the value of the quantity $A(t,\rho)$ at the
$i^\mathrm{th}$ spatial and $j^\mathrm{th}$  temporal node.  The
Neumann boundary condition (\ref{eq:Neumann}) is imposed explicitly,
while the asymptotic boundary conditions (\ref{eq:asymptotic BC 1})
and (\ref{eq:asymptotic BC 2}) are replaced with the Dirichlet
conditions
\begin{equation}
    R(t,\rho_\mathrm{max}) = R_\mathrm{max}, \quad
    V(t,\rho_\mathrm{max}) = 0.
\end{equation}
$R_\mathrm{max}$ is fixed by the \emph{asymptotically flat} initial
data. With this, the former condition effectively gives $dR/d\rho
\sim 1$ for large $\rho$ in our simulations, while the latter leads
to a non-zero value of  $V$ at the origin $\rho=0$.\footnote{This
means that data symmetric about the origin evolves asymmetrically.
Since our computational domain is $\rho \in [0,\rho_\mathrm{max}]$,
the other side of the wormhole (ie. $\rho \in
[-\rho_\mathrm{max},0])$ can be glued continuously using the
junction conditions with a jump in $V$ at $\rho=0$, which represents
the freedom to choose different coordinates on each side of the
join.}  For $\rho_\mathrm{max}$ sufficiently large compared to
$t_\mathrm{max}$, the simulation results are insensitive to the
position of the outer boundary.  In practice, our simulations are
performed with $R_\mathrm{max} \sim \rho_\mathrm{max} \ge 200$.

The two types of finite differencing schemes used to approximate
derivatives are summarized in Table \ref{tab:schemes}. When applied
to linear parabolic systems, the simple Euler (SE) method is known
to be conditionally stable when $\delta t / \delta \rho^2 \lesssim
1$.  Conversely, the modified Dufort-Frankel (MDF) method is known
to be unconditionally stable.  Via direct experimentation, we have
found that these conclusions also hold for the nonlinear problem
given by equations (\ref{eq:flow equation 2}) and
(\ref{eq:constraint equation}), which would na\"{\i}vely suggest
that the MDF method is vastly superior to the SE method due to its
excellent stability.
\begin{table}
\begin{tabular}{c|cc}
    \hline  & Simple Euler (SE) & Modified Dufort-Frankel (MDF) \\
    \hline $\di_\rho A$ & $(A^j_{i+1}-A^j_{i-1})/(2\,\delta \rho)$ &
    $(A^j_{i+1}-A^j_{i-1})/(2\,\delta\rho)$ \\ $\di^2_\rho R$ in eq.~(\ref{eq:flow equation 2}) & $(R^j_{i+1} -
    2R^j_i + R^j_{i-1})/{\delta \rho^2}$ & $ (R^j_{i+1} - R^{j+1}_i -
    R^{j-1}_i + R^j_{i-1})/{\delta \rho^2}$
    \\ $\di^2_\rho R$ in eq.~(\ref{eq:constraint equation}) & $(R^j_{i+1} -
    2R^j_i + R^j_{i-1})/{\delta \rho^2}$ & $(R^j_{i+1} -
    2R^j_i + R^j_{i-1})/{\delta \rho^2}$ \\ $\di_t R$ & $(R^{j+1}_{i}-R^j_{i})/\delta t$ &
    $(R^{j+1}_{i}-R^{j-1}_{i})/(2\,\delta t)$ \\ accuracy &
    $\mathcal{O}(\delta t, \delta\rho^2)$ & $\mathcal{O}(\delta t^2,
    \delta \rho^2)$  \\ \hline
\end{tabular}
\caption{Summary of the simple Euler (SE) and modified
Dufort-Frankel (MDF) finite difference derivative approximations. In
the top line, $A$ stands for either $R$ or $V$.  Notice that for the
MDF method, the second-order spatial derivative of $R$ is handled
differently in the flow (\ref{eq:flow equation 2}) and constraint
(\ref{eq:constraint equation}) equations.}\label{tab:schemes}
\end{table}
However, there is a non-trivial price to be paid for this attractive
feature. One can show that the numerical results obtained with the
MDF approximation actually converge to solutions of
\begin{equation}\label{eq:flow equation 3}
    \di_t R = \di^2_\rho R + \frac{(\di_\rho R)^2}{R} - \frac{1}{R}
    + V \di_\rho R - \left(\frac{\delta t}{\delta \rho} \right)^2 \di_t^2
    R,
\end{equation}
rather than the solutions of the original flow equation
(\ref{eq:flow equation 2}).\footnote{The origin of the anomalous
time derivative on the r.h.s. of (\ref{eq:flow equation 3}) is the
MDF approximation of $\di_\rho^2 R$ in (\ref{eq:flow equation 2}).
Essentially, the MDF prescription approximates $R_i^j$ by its
\emph{temporal} average $(R^{j+1}_i + R^{j-1}_i)/2$.  The use of
nonlocal time data to resolve the spatial derivative gives rise to
the extra $\di_t^2 R$ term.}  In other words, the MDF method solves
a modified version of the original system of PDEs. In order to
minimize the discrepancy between MDF results and the ``true''
solutions of (\ref{eq:flow equation 2}), one must make the ratio
$\epsilon \equiv \delta t/ \delta \rho$ as small as is
computationally feasible.

The numerical results presented in this paper are obtained using the
MDF method with $\epsilon = 10^{-4}$.  For several individual cases,
we have checked that the simulation results are insensitive to the
particular choice of $\delta t$ and $\delta \rho$ provided that
$\epsilon \lesssim 10^{-3}$, and that the MDF and SE schemes give
virtually identical answers for $\epsilon = 10^{-4}$.

\section{Evolution of Morris-Thorne wormhole
geometries}\label{sec:Morris-Thorne}

\begin{figure}
    \begin{center}
    \includegraphics[scale=0.775]{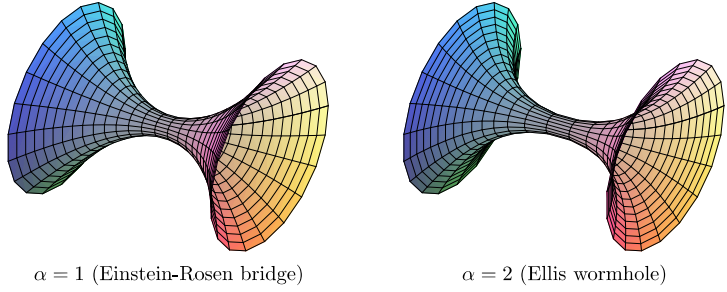}
    \end{center}
    \caption{Embedding diagrams of the Einstein-Rosen and
    Ellis wormholes in flat space (we have set $\theta = \pi/2$ to obtain a
    3-dimensional plot).}\label{fig:embedding 1}
\end{figure}

The Morris-Thorne wormholes \cite{mt-worms} have spatial sections
\begin{equation}
    ds^2 = \frac{1}{f(r)} dr^2 + r^2 d\Omega^2, \quad f(r) = 1 -
    \frac{b(r)}{r}.
\end{equation}
In this section, we will use these metrics as initial data for the
Ricci flow.  At the initial time $t = 0$, this 3-metric can be
transformed into the areal radius gauge (\ref{eq:areal radius
gauge}) by the coordinate transformation $r = R(0,\rho)$ with
\begin{equation}
    \di_\rho R(0,\rho) = \sqrt{f(R(0,\rho))},
\end{equation}
where $R(0,\rho)$ is the initial profile for the areal radius metric
function. We will concentrate on a subset of this class defined by
\begin{equation}
    b(r) = \frac{r_0^\alpha}{r^{\alpha - 1}}, \quad f(r) = 1 -
    \left( \frac{r_0}{r} \right)^\alpha,
\end{equation}
where $r_0$ is a length scale and $\alpha$ is a dimensionless
parameter (a similar family of wormholes was defined in
\cite{Lobo:2005us}). The spatial sections of the Schwarzschild
metric (known as the Einstein-Rosen bridge) correspond to $\alpha =
1$. Another special case is the Ellis wormhole with $\alpha = 2$
\cite{ellis}, which has $R(0,\rho) = \sqrt{\rho^2 + r_0^2}$.
Embedding diagrams of both types of wormholes appear in Figure
\ref{fig:embedding 1}. For Ricci flow of these wormholes, it is
useful to work with dimensionless quantities defined by
\begin{equation}\label{eq:dimensionless}
    \hat\rho = \rho/r_0, \quad \hat{t} = t/r_0^2, \quad \hat{R}
    \rightarrow R/r_0, \quad \hat{V} \rightarrow r_0 V.
\end{equation}
When the hatted-variables are substituted into the flow equations
the $r_0$ length scale drops out of the problem; i.e., our
simulation results are essentially independent of $r_0$ up to
trivial scalings.
\begin{figure}
    \begin{center}
    \includegraphics[scale=0.775]{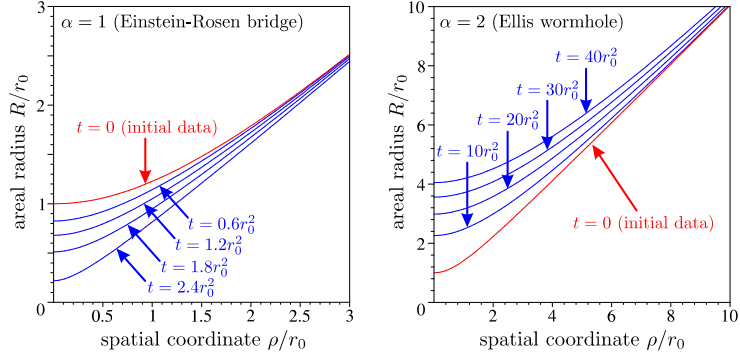}
    \end{center}
    \caption{Evolution of the areal radius function $R(t,\rho)$
    under the Ricci flow for Morris-Thorne wormhole initial data
    corresponding to $\alpha = 1$ and $\alpha = 2$.  Note that our
    boundary conditions ensure that the wormhole throat is always at $\rho = 0$
    for these simulations.}\label{fig:explicit evolution}
\end{figure}

In Figure \ref{fig:explicit evolution}, we show the results of our
simulations for the evolution of the areal radius function
$R(t,\rho)$ for the Einstein-Rosen and Ellis cases. One can see in
Figure \ref{fig:embedding 1} that the initial data for the two
scenarios are visually quite similar. However, the evolution of $R$
is entirely different: in the Ellis case the areal radius tends to
diverge for all $\rho$ as $t \rightarrow \infty$, while for the
Einstein-Rosen bridge $R$ is everywhere decreasing and the wormhole
throat pinches off in finite time.  Due to computing limitations, we
cannot determine the ultimate fate of the Ellis initial data, but it
appears to be a warped cylinder with a monotonically increasing
radius.

\begin{figure}
    \begin{center}
    \includegraphics[scale=0.775]{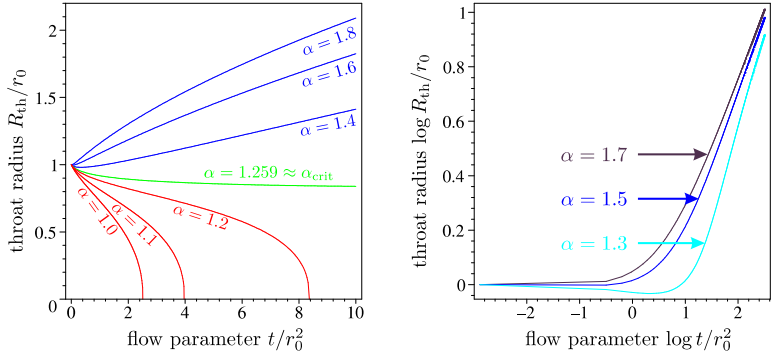}
    \end{center}
    \caption{Ricci flow evolution of the throat radius $R_\mathrm{th}(t) = R(t,0)$
    for various Morris-Thorne wormhole initial data.}
    \label{fig:throat evolution}
\end{figure}

In Figure \ref{fig:throat evolution}, we plot the evolution of the
throat radius $R_\mathrm{th}(t) = R(t,0)$ for various values of
$\alpha$.  We find that the wormhole throat pinches off for initial
data with $\alpha$ less than a critical value
$\alpha_\mathrm{crit}$.  For cases with $\alpha >
\alpha_\mathrm{crit}$, the throat approaches an ever-expanding state
after a brief initial period of contraction. Numerically, we have
determined the value of the critical parameter to be approximately
1.259.  For $\alpha > \alpha_\mathrm{crit}$, the throat radius
appears to grow like $\sqrt{t}$ at late times.

\section{Evolution of ``bubble'' geometries}\label{sec:bubbles}

\begin{figure}
    \begin{center}
    \includegraphics[scale=0.775]{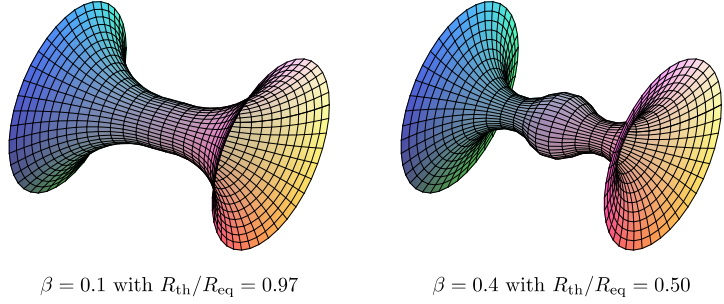}
    \end{center}
    \caption{Embedding diagrams of 3-geometries representing a bubble
    connecting two asymptotically flat regions}\label{fig:embedding 2}
\end{figure}

A geometrically interesting question is: What does the Ricci flow
evolution of a bubble attached to asymptotically flat space look
like? In the areal radius gauge, initial data for this kind of
problem would roughly be
\begin{equation}\label{eq:bubble data 1}
    R(0,\rho) \sim r_0
    \cases{ \sin \rho / r_0 & for $0 \le \rho \lesssim \rho_0$ \\
    \rho/r_0 & for $\rho \gtrsim \rho_0$ },
\end{equation}
where $\rho_0 < \pi r_0$ is some transition radius.  Notice that
this initial data has $R = 0$ and $\di_\rho R \ne 0$ at $\rho = 0$,
so it violates the boundary conditions that our code was designed to
deal with.  We therefore consider the related problem of a bubble
attached to two asymptotically flat regions, and leave initial data
of the form (\ref{eq:bubble data 1}) for future work. We take
initial data for the current problem to be
\begin{equation}
    R(0,\rho) =
    \cases{ r_0 (1-\rho^2/r_0^2) & for $|\rho| < \rho_0$ \\
    \sqrt{(|\rho| - \rho_1)^2 + \rho_2^2}  & for $|\rho| \ge \rho_0$
    }.
\end{equation}
Here, $\rho_1$ and $\rho_2$ are selected such that both $R$ and
$\di_\rho R$ are continuous at the transition points $\rho = \pm
\rho_0$.  Defining the dimensionless parameter $\beta \equiv
\rho_0/r_0$, we show embedding diagrams for two examples of this
type of initial data in Figure \ref{fig:embedding 2}.  Note that
there are two wormhole throats (local minima of $R$) in this case at
$\rho = \pm \rho_1$, while there is a local maxima of $R$ at $\rho =
0$. This maxima corresponds to the equator of our bubble, and it is
easy to show that the ratio of the throat radii to the equator
radius is
\begin{equation}
    {R_\mathrm{th}}/{R_\mathrm{eq}} = \rho_2/r_0 = (1-\beta^2)
    \sqrt{1-4\beta^2}.
\end{equation}
Hence, in order to have sensible initial data with $R > 0$ we are
obliged to take $0 \le \beta \le 1/2$.

Making use of the dimensionless quantities (\ref{eq:dimensionless}),
we have simulated the Ricci flow evolution of various types of
bubble geometries.  As in \S\ref{sec:Morris-Thorne} above, we take
$R$ to be an even function of $\rho$ and impose the boundary
condition that $\di_\rho R = 0$ for $\rho = 0$.  We find that for
$R_\mathrm{th}/R_\mathrm{eq}$ initially small, the bubbles disappear
and the geometry ``pinches-off'' into a pair of disjoint
asymptotically flat regions. For $R_\mathrm{th}/R_\mathrm{eq}$
initially close to unity, the bubbles also disappear and the areal
radius blows up, somewhat similar to the $\alpha >
\alpha_\mathrm{crit}$ cases in \S\ref{sec:Morris-Thorne}. Explicit
examples of these behaviours are shown in Figure \ref{fig:bubble
evolution}.
\begin{figure}
    \begin{center}
    \includegraphics[scale=0.775]{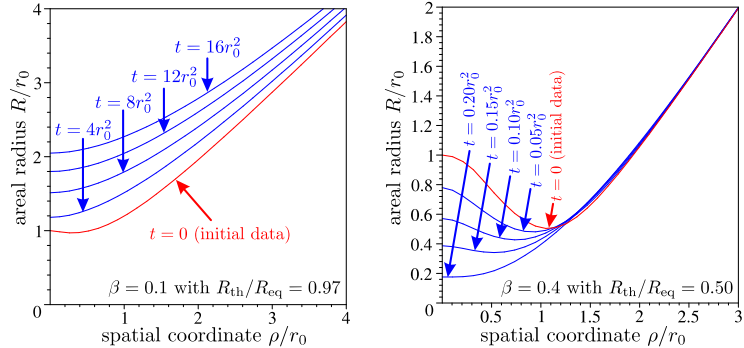}
    \end{center}
    \caption{The evolution of two bubble geometries under the Ricci flow.}
    \label{fig:bubble evolution}
\end{figure}

\section{Conclusions}

We have observed that the Ricci flow of certain  families  of
rotationally symmetric wormhole geometries exhibit a form of
critical behaviour:  wormhole throats pinch off or expand forever
depending on the values of initial data parameters. The expanding
solutions are perhaps surprising at first sight, but are
nevertheless understandable in that they appear to converge to an
infinite volume cylindrical attractor.

This work  complements both the analytic work on the flow of
asymptotically flat geometries \cite{ow-asympflat} and the numerical
work on corsetted  3--spheres  \cite{gi}.  It would be of interest
to see if the neck pinching seen in our  simulations of wormhole
throats may be modeled by the Bryant steady solitons, as has been
observed for rotationally symmetric geometries on $S^3$ \cite{gi2}.

The methods we have used have applicability to a variety of
situations for relevance to string theory, where the renormalization
group flows give matter coupled modifications of Ricci flow. The
interest here is in discovering new fixed points of physical
relevance. For example, in special cases with compact target spaces
and  a tachyon field, there is evidence that properties of the
tachyon potential are linked to the existence of non-trivial fixed
points \cite{gs-tachyon}. It  is of interest to extend such cases to
non-compact geometries with gauge or 2-form fields, and consider
questions such as the possibility of stabilization of  wormholes
geometries.
\medskip

\noindent{\bf Acknowledgements} We would like to thank Eric Woolgar
and Jack Gegenberg for discussions, David Hobill for suggesting that
we try the Dufort-Frankel method, and David Garfinkle for comments
on the manuscript. This work was supported by NSERC.

\end{document}